\documentclass[12pt,aps,prb,preprint]{revtex4}   % style for Physical Review B and AJP are similar
\usepackage{amsmath}    % need for subequations
\usepackage{graphicx}   % for figures
\begin{document}

\title{A different derivation of the acceleration of a particle in a non-inertial reference frame}
\author{Eli Lansey}
\affiliation{Department of Physics, City College and The Graduate Center of the City University of New York, New York, NY 10031}
\email{elansey@gc.cuny.edu}
\date{\today}

\begin{abstract}
%A derivation of the acceleration of a particle in a non-inertial reference frame, including the Coriolis effect, is a standard part of most undergraduate  and graduate-level courses in mechanics.  A new approach to this derivation using the geometrical meaning of complex variables is developed.  This method allows easier comprehension of various cross-product terms typically appearing in these expressions.
A new derivation of the acceleration of a particle in a non-inertial reference frame, including the Coriolis effect, is developed using the geometry of complex numbers in the complex plane.
\end{abstract}

\maketitle

\section{Introduction}
Many derivations of the acceleration of a particle in a non-inertial reference frame are purely mathematical.  That is, although fundamentally based on the physical properties of the rotating system, the proofs quickly move into mathematically rigorous $\epsilon_{ijk}$ notation,~\cite{goldstein} or manipulations of physically unintuitive Lagrangians.~\cite{landau}\@  Furthermore, attempts at deriving these results in undergraduate texts~\cite{undergrad} involve messy, unclear calculations. % involving, among other issues, derivatives of unit vectors

In all of these methods, the {\it physical meaning} of result is almost secondary to the mathematics of the derivation.  That is, one needs to pick through the resultant vector product equations carefully to understand the physical meaning of each term in the final expression.  Furthermore, these derivations use, by and large, single-use tools, which have few applications outside of this unique problem, and do not introduce any broader mathematical or physical tools to the reader.

In this paper I develop a different method for deriving the acceleration of a particle in a non-inertial reference frame using complex variables.  As the graphical meaning of complex multiplication involves rotations in the complex plane, it is a natural tool for studying rotations of any kind, not just non-inertial reference frames.  Familiarity with complex numbers in general, and an intuition into their meaning and graphical properties, is essential to students as they progress into electromagnetism, quantum mechanics and other areas of physics.  Furthermore, each step of this derivation is primarily motivated by physical arguments, and the physical meaning of the resulting expression is significantly easier to understand.

\newcommand{\rp}{{\vec{r}\,'}}
\newcommand{\oper}{{\tilde{\Theta}}}

\section{Geometrical setup and the rotation operator}
\begin{figure}%[htbp]
	\centering
		\includegraphics[width=0.50\textwidth]{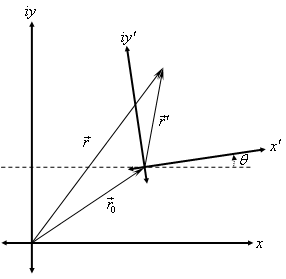}
	\caption{The 2D geometrical setup of a translating and rotating reference frame}
	\label{fig:fig1}
\end{figure}

%\begin{figure}
%  % Requires \usepackage{graphicx}
%  \includegraphics[width=]{}\\
%  \caption{}\label{}
%\end{figure}

Consider a particle moving in a (primed) reference frame moving with respect to a fixed reference frame (see Fig.~\ref{fig:fig1}, ignore the $i$'s for the moment).  The vector position $\vec{r}$ of the particle is
\begin{equation}
\label{eq:nomap}
\vec{r} = \vec{r}_0+\rp ,
\end{equation}
where $\vec{r}_0$ is the position of the origin of the moving reference frame and $\rp$ is the position of the particle relative to the moving reference frame.  In a case of translational motion, where the elements $\rp$ are given with respect to the basis of the fixed system (i.e. $\vec{r}_0=x_0\,\hat{i}+y_0\,\hat{j}$ and $\rp=x'\,\hat{i}+y'\,\hat{j}$ in 2D), the interpretation of this formula is straightforward vector addition:
\begin{equation}
\vec{r} = (x_0+x\,')\hat{i}+(y_0+y\,')\hat{j} .
\end{equation}

However, in the more common case where $\rp$ is given with respect to a {\it rotating} basis (i.e. $\rp=x\,'\hat{i}\,'+y\,'\hat{j}\,'$), one needs to know how the rotating basis vectors map to the unrotated ones.  To that end, we introduce an operator $\oper$ that defines the rotation of the system, and maps a vector given with respect to the moving basis to a vector given with respect to the fixed basis, thus allowing simple vector addition.

In a non-rotated and non-rotating frame, for example, this operator is simply the identity operator $\oper=\tilde{I}$.  Another familiar rotation operator (in 2D) is the matrix of rotation
\begin{equation*}
\oper\,=
\begin{pmatrix}
\,\,\,\cos \theta \hfill & \,\,\sin \theta \\
-\sin \theta \hfill & \,\,\cos \theta  \hfill
\end{pmatrix}.
\end{equation*}
\noindent In general, however, this operator need not be a matrix.  Additionally, $\oper$ can, and usually will, depend explicitly  on time.  Furthermore, since $\oper$ is an operator, it is important to note that $\oper\vec{r}\neq\vec{r}\oper$; the left hand side is a vector, while the right hand side is still an operator.

\section{Vector position as a complex vector}
We can now write the time-dependent vector position of the particle as
\begin{equation}
\label{eq:rmap}
\vec{r}(t) = \vec{r}_0(t)+\oper(t)\rp(t) .
\end{equation}
Assuming a fixed axis of rotation, we can study the effects of the rotation in a 2D system without any loss of generality.
We now let a particle's vector position $\vec{r}(t)$ in 2D Cartesian coordinates be expressed as a complex vector, such that $x(t)=\text{Re}[\vec{r}(t)]$ and $y(t)=\text{Im}[\vec{r}(t)]$:
\begin{equation}
\vec{r}(t)=x(t)+iy(t).
\end{equation}
The other vector quantities $\vec{r}_{0}(t)$ and $\rp(t)$ should likewise be considered complex vectors.

To write the operator $\oper$ in this notation, we must recall that multiplying a complex number $z=re^{i\theta}$~-- graphically, a vector of length $r$ making angle $\theta$ with respect to the $x$-axis~-- by $z_1=r_1e^{i\theta_1}$ gives $z_1z=(r_1r)e^{i(\theta+\theta_1)}$ .  This can be understood graphically as a scaling of $r$ by $r_1$ and a rotation of the direction of $z$ by the angle $\theta_1$.~\cite{needham}  So, applying a rotation to a complex position vector $\rp$ is as simple as multiplying it by the unit-magnitude complex number $e^{i\theta(t)}$, where $\theta$ is the angle of rotation, namely the angle between the $x$ and $x\,'$ axes (see Fig.~\ref{fig:fig1}, and this time don't ignore the $i$'s).

\newcommand{\operet}{{\tilde{e}^{i\theta(t)}}}
\newcommand{\opere}{{\tilde{e}^{i\theta}}}

Thus we have
\begin{equation}
\label{eq:operdef}
\oper(t)=\operet,
\end{equation}
and can write the position of the particle as
\begin{equation}
\label{eq:pos}
\vec{r}(t)=\vec{r}_0(t)+\operet\rp(t).
\end{equation}
Please note that $e$ is wearing a tilde to remind us of its job as the rotation operator.

\section{Velocity}
\label{velocity}
If we wish to find the velocity of the particle at any point time, we differentiate $\vec{r}$ with respect to time.  To keep things uncluttered, the explicit designation of functions of time will be dropped.
\begin{equation}
\label{eq:v1}
\vec{v}\equiv\frac{d\vec{r}}{dt}=\frac{d}{dt}\bigr[\vec{r}_0+\opere\rp\bigl]=\vec{v}_0+\frac{d}{dt}\bigr[\opere\rp\bigl].
\end{equation}
Here $\vec{v}_0\equiv d\vec{r}_0/dt$, which is the translational velocity of the origin of the moving and rotating system.  Because of the operator in the second derivative term of Eq.~\ref{eq:v1}, we will expand this derivative assuming non-commutative multiplication:
\begin{equation}
\label{eq:v2}
\frac{d}{dt}\bigr[\opere\rp\bigl]=\opere\frac{d}{dt}\bigr[\rp\bigl]+\frac{d}{dt}\bigr[\opere\bigl]\rp=\opere\vec{v}\,'+\opere \widetilde{i\omega} \rp=\opere\bigr(\vec{v}\,'+\widetilde{i\omega} \rp\bigl).
\end{equation}
Here $\vec{v}\,'\equiv d\rp/dt$, i.e. the translational velocity of the particle within the rotating system, and $\omega\equiv d\theta/dt$, i.e. the angular velocity of the rotating coordinate system.  Thus we have
\newcommand{\vp}{{\vec{v}\,'}}
\begin{equation}
\label{eq:v}
\vec{v}=\vec{v}_0+\opere \bigr(\vp+\widetilde{i\omega}\rp\bigl).
\end{equation}

Let's make sense of each component.  As mentioned earlier, $\vec{v}_0$ is the translational velocity of the moving system.  Additionally, $\vp$ is the translational velocity of the particle within the moving system, which the rotation operator maps to the appropriate basis.

The last term is more difficult.  However, due to complex notation, $i=e^{i\frac{\pi}{2}}$.  In other words, $i$ represents a rotation of 90 degrees, so that $a\bot ia$.  However, let's not forget that $\widetilde{i\omega}$ is still functioning as an operator, as it came from the derivative of the rotation operator.  However, since $\omega$ is a scalar, and $\oper\omega=\omega\oper$, we can write the operator $\widetilde{i\omega}$ as $\omega\tilde{i}$.

Thus, we look at $\omega\tilde{i}\rp$ as a vector with magnitude $\omega|\rp|$ pointing perpendicular to $\rp$.  We recognize this vector as the tangential velocity of a particle traveling in a circle of radius $|\rp|$ with angular velocity $\omega$.  Thus, the last term is understood as well.  This is just the contribution of the tangential velocity of the particle coming from its rotation about the primed origin.

Given this discussion, we can now write Eq.~\ref{eq:v}
\begin{equation}
\label{eq:vv}
\vec{v}=\vec{v}_0+\opere \bigr(\vp+\omega\tilde{i}\rp\bigl).
\end{equation}

\section{Acceleration and Coriolis Effect}
To find the acceleration of the particle, we can apply the same method as in Section~\ref{velocity}, and differentiate the velocity as given by Eq.~\ref{eq:vv} with respect to time:
\begin{equation}
\label{a1}
\vec{a}\equiv\frac{d\vec{v}}{dt}=\frac{d}{dt}\bigr[\vec{v}_0+\opere (\vp+\omega\tilde{i}\rp)\bigl]=\vec{a}_0+\frac{d}{dt}\bigr[\opere (\vp+\omega\tilde{i}\rp)\bigl].
\end{equation}
Here $\vec{a}_0\equiv d\vec{v}_0/dt$, which is the translational acceleration of the origin of the moving system.  Again, the second term needs careful attention:
\begin{equation}
\label{a2}
\frac{d}{dt}\bigr[\opere (\vp+\omega\tilde{i}\rp)\bigl]=\opere \biggl\{\omega\tilde{i}(\vp+\omega\tilde{i}\rp)+\vec{a}\,'+\frac{d}{dt}\bigr[\omega\tilde{i}\rp\bigl]\biggr\}.
\end{equation}
Here $\vec{a}\,'\equiv d\vp/dt$, i.e. the translational velocity of the particle within the rotating system.  The final term $\frac{d}{dt}[\omega\tilde{i}\rp]=\omega\tilde{i}\vp+\alpha\tilde{i}\rp$, where $\alpha\equiv d\omega/dt$, i.e. the angular acceleration of the rotating system.  Thus, the complete expression for acceleration is:
\begin{equation}
\label{eq:acc}
\vec{a}=\vec{a}_0+\opere\bigl(\vec{a}\,'+\alpha\tilde{i}\rp+\omega\tilde{i}\omega\tilde{i}\rp+2\omega\tilde{i}\vp\bigl).
\end{equation}

Now we need to understand the various elements.  As mentioned before, $\vec{a}_0$ is the acceleration of the moving system, and $\vec{a}\,'$ is the translational acceleration of the particle within the rotating system. The other three elements require further consideration.  Keeping in mind the earlier discussion about the operator $\tilde{i}$, we realize that $\alpha\tilde{i}\rp$ is the tangential acceleration from the angularly accelerating coordinate system, also called the transverse acceleration.

Additionally, $\omega\tilde{i}\omega\tilde{i}\rp=\omega^2\tilde{i}^2\rp$.  Now, recall that $\tilde{i}$ represents a rotation of 90 degrees.  So, applying that rotation twice gives a rotation of 180 degrees.  This is seen explicitly by $ii=e^{i\frac{\pi}{2}}e^{i\frac{\pi}{2}}=e^{i\pi}$.  Thus, $\omega^2\tilde{i}^2\rp$ is a vector pointing along $\rp$ but with opposite sign, with magnitude $\omega^2|\rp|$.  We recognize this as the expression for centripetal acceleration.

However, $2\omega\tilde{i}\vp$ is more complicated.  We see that it's similar in structure to tangential velocity, being that it has a $\omega\tilde{i}$ operator, so we know it's a vector pointing perpendicular to the velocity of the particle in the moving system.  Such an acceleration will not cause any change in the speed of the particle, and as such, in the absence of any other accelerations, will pull the particle into a circular path.   That is, if the particle is moving right, the acceleration points up, if the particle is moving up, the acceleration points left, and so on.

There is a similar expression for the acceleration vector of a charged particle moving in a magnetic field.  Translated into our notation, it is given as $-\frac{q}{m}B\tilde{i}\vec{v}$, where the magnetic field $B$ points out of the surface, and we know that such a particle travels in a circular path, as well.  In certain circumstances, then, this can also be a source of centripetal acceleration.

This particular spirally sort of acceleration is called the Coriolis acceleration, and is what causes hurricanes to spiral.  If you look down the axis of the spinning earth from the north pole, and send air moving up (north) - it will be accelerated left (west), then down (south), then right (east), then up (north, again) - and a circular hurricane will form.  Looking from the south pole will switch the direction of $\omega$ and cause things to spiral in the opposite direction.

\section{Generalization to 3 Dimensions}
Now that all the components of Eq.~\ref{eq:vv} and Eq.~\ref{eq:acc} are understood, these expressions can easily be generalized for 3D systems as well.  Note that in both expressions we still have the original rotation operator $\opere$ in its original position, unchanged by any of our manipulations.  This can be re-generalized back to an arbitrary 3 dimensional operator $\oper$.  However, the operator $\omega\tilde{i}$ needs to be understood in 3D.

First, define the vector $\vec{\omega}$  as a vector with magnitude $\omega$ pointing along the axis of rotation (following the ``right hand rule'').  Recalling that $\omega\tilde{i}\rp$ is the tangential velocity, we want to find an expression for tangential velocity in 3D around our arbitrary axis of rotation.  To that end, we need to know the distance between the axis of rotation and the particle, and multiply it by the angular velocity.  This is simply $\vec{\omega}\times\rp$.

Thus we have the operator $\omega\tilde{i}\mapsto\vec{\omega}\times$.  Noting that we can replace the symbol $\omega$ with anything we want, we have a simpler operator relationship
\begin{equation}
\label{eq:imap}
\tilde{i}\mapsto\vec{}\times.
\end{equation}

So, using Eq.~\ref{eq:imap} in Eq.~\ref{eq:vv} and Eq.~\ref{eq:acc}, we can now write a general expression for the velocity and acceleration of a particle in a rotating and translating coordinate system:
\begin{subequations}
\label{eq:final}
\begin{align}
\label{eq:vfinal}\vec{v}&=\vec{v}_0+\oper \bigr(\vp+\vec{\omega}\times\rp\bigl) \\
\label{eq:afinal}\vec{a}&=\vec{a}_0+\oper\bigl(\vec{a}\,'+\vec{\alpha}\times\rp+\vec{\omega}\times(\vec{\omega}\times\rp)+2\vec{\omega}\times\vp\bigl) .
\end{align}
\end{subequations}
Here $\vec{\alpha}\equiv d\vec{\omega}/dt$, i.e. the vector angular acceleration of the system, which in this case points along $\vec{\omega}$.  Thus $\vec{\alpha}\times\rp$ is the general expression for transverse acceleration, $\vec{\omega}\times(\vec{\omega}\times\rp)$ is the general expression for centripetal acceleration, and $2\vec{\omega}\times\vp$ is the general expression for Coriolis acceleration.

To find the velocity and acceleration of the particle within the rotating system, solve Eqs.~\ref{eq:final} for $\vp$ and $\vec{a}\,'$:
\begin{subequations}
\label{eq:pfinal}
\begin{align}
\label{eq:vpfinal}\vp &= \oper^{-1}(\vec{v}-\vec{v}_0) -\vec{\omega}\times\rp \\
\label{eq:apfinal}\vec{a}\,'&=\oper^{-1}(\vec{a}-\vec{a}_0)- \vec{\alpha}\times\rp - \vec{\omega}\times(\vec{\omega}\times\rp) - 2\vec{\omega}\times\vp .
\end{align}
\end{subequations}
Here $\oper^{-1}$ is defined such that $\oper^{-1}\oper\equiv\tilde{I}$, and is the inverse rotational transformation.  In the complex space described earlier, for example, $\oper^{-1}=\tilde{e}^{-i\theta}$.

From Eq.~\ref{eq:apfinal} we see that from the motion of a particle we can determine if we're in a rotating system by measuring the net acceleration of the particle.  For example, on Earth's equator $\vec{\alpha}\approx 0$, $\vec{a}-\vec{a}_0=0$, and $\vec{\omega}\times(\vec{\omega}\times\rp)=0$.  Thus, Eq.~\ref{eq:apfinal} becomes $\vec{a}\,' = - 2\vec{\omega}\times\vp$.  Directing the velocity of the particle along the equator gives a non-zero value for this acceleration.  Then, if there are no direct forces on a moving particle we'd expect the particle to follow a straight path.  Yet, due to the Coriolis acceleration, the particle's path will bend northwards, and one can see direct evidence of the rotation of the Earth.

\begin{acknowledgments}
I wish to thank Profs. J.I. Gersten and T.H. Boyer for their helpful suggestions on an earlier version of the manuscript.
\end{acknowledgments}

\end{document}